\documentclass[journal,twoside]{ieeetran}
\usepackage{amsmath,amssymb,amsfonts}
\usepackage{algorithmic}
\usepackage{graphicx}
\usepackage{textcomp}
\usepackage{float}

\usepackage{booktabs}
\begin{document}

\title{Harnessing Smartwatch Microphone Sensors for Cough Detection and Classification
}
 
\author{Pranay Jaiswal,  Haroon R Lone, \IEEEmembership{Member, IEEE}
\thanks{This paragraph of the first footnote will contain the date on 
which you submitted your paper for review. It will also contain support 
information, including sponsor and financial support acknowledgment.}
\thanks{Corresponding author: Haroon R Lone (e-mail: haroon@iiserb.ac.in).}
\thanks{Pranay Jaiswal and Haroon R Lone are with the Department of EECS, Indian Institute of Science Educaton and Research Bhopal, Madhya Pradesh, 462066, India (e-mail: pranay18@iiserb.ac.in; haroon@iiserb.ac.in).}
 }

\maketitle

\begin{abstract}
This study investigates the potential of using smartwatches with built-in microphone sensors for monitoring coughs and detecting various cough types. We conducted a study involving 32 participants and collected 9 hours of audio data in a controlled manner. Afterward, we processed this data using a structured approach, resulting in 223 positive cough samples. We further improved the dataset through augmentation techniques and employed a specialized 1D CNN model. This model achieved an impressive accuracy rate of 98.49\% while non-walking and 98.2\% while walking, showing smartwatches can detect cough. Moreover, our research successfully identified four distinct types of coughs using clustering techniques.
\end{abstract}


\section{Introduction}
Coughing is a fundamental physiological response, often used as a crucial indicator of underlying health conditions~\cite{chang2006physiology}. For individuals with lung-related ailments, increased cough frequency can signal the onset of an episode or a decline in health. Consistently monitoring cough patterns allows us to observe the health status of those at risk of lung diseases, those currently affected, or those with existing conditions~\cite{alqudaihi2021cough}.

Beyond chronic lung diseases, cough monitoring is pivotal in identifying and tracking illnesses such as COVID-19, where coughing is a prominent symptom \cite{9208795, pahar2021covid}. Regularly tracking coughs in healthy individuals establishes a baseline for respiratory health, aiding in the early detection of health issues or emerging infections. This monitoring empowers healthcare professionals and individuals to gain insights into respiratory health, track disease progression, and facilitate targeted interventions during outbreaks, improving public health measures.

Smartphones have proven valuable gadgets in detecting coughs in recent years\cite{cesnakova2019continuous}. They are well-suited for this task with their built-in microphones and impressive computing abilities. Scientists and app developers have capitalized on these features to create mobile applications that utilize advanced audio analysis methods to differentiate between cough sounds and ambient noise\cite{9208795, pahar2021covid}. However, smartphones encounter certain obstacles in accurately detecting coughs. These include microphone quality variances, microphone placement discrepancies across different smartphone models, and the possibility of obstruction when phones are kept in pockets or bags. Consequently, these factors can impede the consistent and reliable analysis of cough sounds.

Wearable devices with sensitive microphones have emerged as a practical solution for audio data collection in sound-based monitoring. Analyzing coughing frequencies, for instance, can provide valuable insights into respiratory health, aiding in the early detection of respiratory infections or conditions. Sound-based monitoring can be further extended to encompass the detection of breathing patterns, snoring, or other specific sounds associated with various health conditions or behaviors\cite{liaqat2021coughwatch, cesnakova2019continuous, vhaduri2020nocturnal, 9208795}.

This paper presents the results of a user study to evaluate the efficacy of a smartwatch's cough detection. We developed a smartwatch app for data collection and conducted a user study during which participants performed different activities. Additionally, we have developed a highly accurate algorithm capable of identifying coughs and have implemented advanced clustering techniques to differentiate between different types of coughs. Our study yielded impressive results, achieving a remarkable accuracy rate of 98.49$\%$ in detecting coughing events. Notably, our system demonstrated the capability to distinguish between four distinct types of cough with high precision. Through this work, we contribute to the evolving field of sound-based health monitoring, offering a valuable tool for healthcare professionals and individuals to maintain and improve respiratory health. 

 The paper highlights the following contributions: 
\begin{enumerate}
\item  Design and conduct a user study to collect the cough data with a smartwatch 
\item  Develop algorithms for cough detection. 
\item  Public release of the smartwatch app and the collected data set for wider accessibility~\cite{our-dataset}
\end{enumerate} 
\begin{figure*}[!t]
     \centering
     \includegraphics[width=1\linewidth]{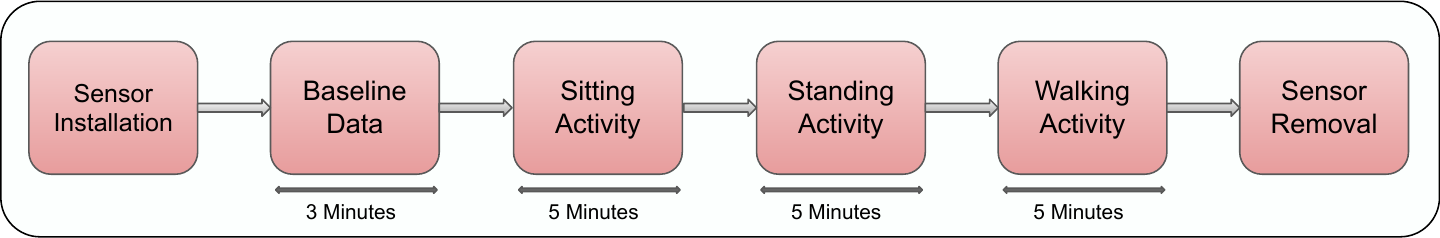}
     \caption{Study Design}
     \label{study design}
 \end{figure*}
\section{Related Work}
Ubiquitous consumer-grade devices such as smartphones and wearables (e.g., smart watches) are being explored for health monitoring applications. In this section, we discuss recent works in which smartphones and smartwatches were used for cough detection. 

\noindent{\bf Cough detection with smartphones:}
Sudip Vhaduri~\cite{vhaduri2020nocturnal} detected coughs and snores using smartphone microphones and various modeling schemes. He analyzed three datasets and different types of nocturnal noises. Later, he employed Mel-frequency cepstral coefficient (MFCC) features and classification techniques. Results showed that the personalized random forest (RF) model yielded an average accuracy of 0.96, F1 score of 0.96, and AUC-ROC of 0.98. 

Laguarta et al.~\cite{9208795} used forced-cough smartphone recordings to discriminate COVID-19 subjects. They built a dataset with 5,320 subjects, developed an AI framework leveraging acoustic biomarker features, and employed a Convolutional Neural Network (CNN) architecture. The CNN achieved a COVID-19 sensitivity of 98.5$\%$ with a specificity of 94.2$\%$. Similarly, \cite{pahar2021covid} used smartphone recordings and a machine learning classifier for COVID-19 cough classification. The Resnet50 classifier performed exceptionally well, with an AUC of 0.98 in identifying COVID-19-positive coughs.

Also, Cesnakova et al.~\cite{cesnakova2019continuous} detected cough from recorded smartphone audio. They developed a smartphone application for sound collection and a mathematical model for cough identification. The model achieved promising results, with a cough recognition sensitivity of 90$\%$ at a 99.5$\%$ specificity preset and 75$\%$ at a 99.9$\%$ specificity preset when tested on publicly available data.

\noindent{\bf Cough detection with smart watches:}
Liaqat et al.~\cite{liaqat2021coughwatch} introduced ``CoughWatch'', a cough detection system designed for smartwatches. This system utilizes both audio and movement data to identify coughs in real-world settings. The authors collected a substantial dataset, including 4,225 hours of sensor and audio data from real-world scenarios and 12 hours of in-lab audio data. CoughWatch achieved impressive results on in-the-wild data, with a precision rate of 82$\%$ and a recall rate of 55$\%$. Additionally, CoughWatch improved precision by up to 15.5 percentage points compared to an audio-only model by incorporating gyroscope and accelerometer data. 

Our research centers around evaluating the capability of smartwatches to accurately detect and classify coughs amidst various types of background noises during different activities.

 \section{Data collection}
\noindent{\bf Smartwatch App:} We created a custom Android application using Android Studio and the Kotlin programming language to gather audio data from a Samsung smartwatch, specifically the Galaxy Watch 4. Within the application, we integrated a Microphone module, which is responsible for capturing audio data. This module allows us to record various sounds and voices, enabling the identification of coughing incidents, speech patterns, and other sound-related data pertinent to health monitoring. Furthermore, we leveraged Amazon Web Services (AWS) S3 for securely storing the acquired audio data files in the cloud.

\noindent{\bf User study:}
The study's design is illustrated in Figure~\ref{study design}. Initially, a smartwatch is installed on the participants'  non-dominant hand. Once installed, data collection by the smartwatch commences. The smartwatch stores this data in MP3 files. A three-minute data collection period is initiated to establish a baseline while participants remain seated without any movement or coughing.

Following this baseline period, participants engage in activities that mimic sitting, standing, and walking, each lasting for five minutes. During the sitting activity, participants may perform tasks such as working on a laptop, writing, or using a mobile phone. In the standing activity, participants engage in actions like drinking water or having phone conversations. Finally, participants simulate walking for five minutes. At various intervals during these activities, participants are instructed to mimic coughing sounds.

At the conclusion of the study session, the smartwatch is removed, and the collected data is transferred to the cloud. The entire study session for each participant typically lasts between 25 to 30 minutes.
 
\noindent{\bf Dataset:} Participants for the study were recruited through email invitations, and individuals who expressed interest indicated their availability via a Google form. Subsequently, these participants visited our research laboratory to take part in the study. Before the study began, all participants provided signed consent forms as a prerequisite.

We enrolled thirty-two student participants in the age range of 20 to 28, with an emphasis on ensuring diversity within this particular age bracket. Our recruitment process was designed to create a representative sample of participants. The data collection phase spanned 28 days, during which all participants actively engaged in study-related activities.

Throughout this period, we gathered a total of 9 hours of audio data from the smartwatch. The adopted sampling frequency for this data collection was 16 kHz.

\section{Methodology}

\begin{figure*}[]
     \centering
     \includegraphics[width=1\linewidth]{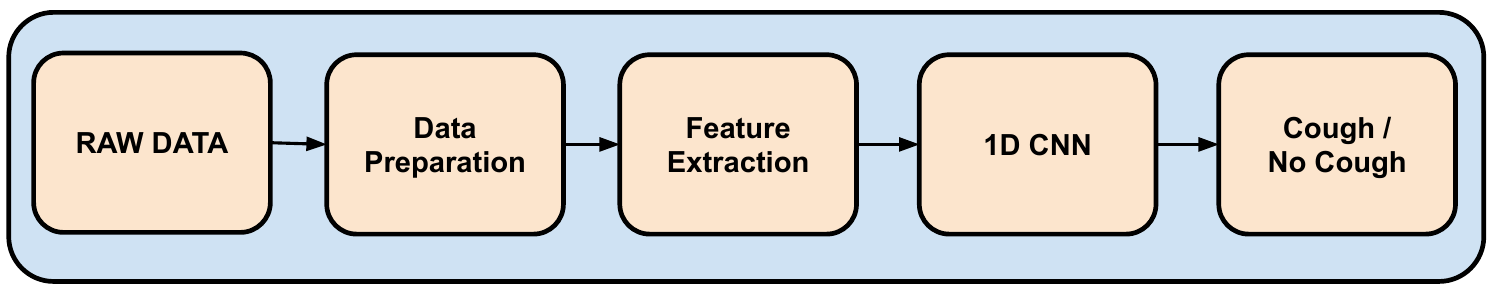}
     \caption{Steps in cough detection}
     \label{cough-steps}
 \end{figure*}
 
Cough detection and the subsequent classification into distinct clusters involve a structured pipeline consisting of three fundamental steps: Data preparation, feature extraction, and classification. The following section provides a detailed description of each of these steps. Figure~\ref{cough-steps} provides an overview of the steps. 

\subsection{Data preparation}

 A total of 9 hours of audio data was collected from the smartwatch. We followed \cite{liaqat2021coughwatch} while preparing the data for analysis. Following is a description of the steps. 

\subsubsection{Manual cough data labeling and extraction}
\begin{figure}[h]
     \centering
\includegraphics[width=1\linewidth]{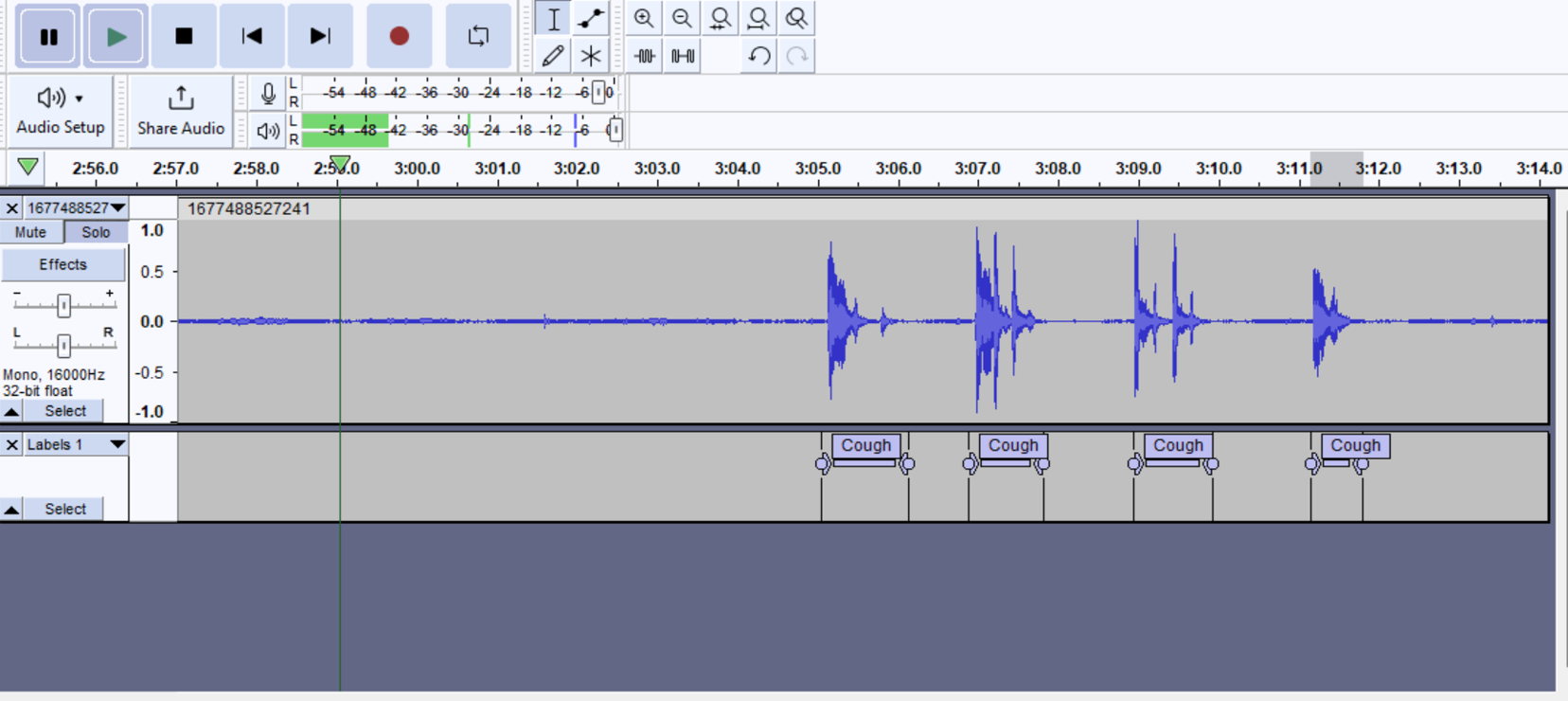}
     \caption{An example of cough file extraction in Audacity Software.}
     \label{auda}
\end{figure}
 
The recorded audio files in WAV format were opened in Audacity software for manual cough labeling, as shown in Figure~\ref{auda}. The audio file's waveform was examined, and the cough sound area was zoomed in for better visualization. The cough sounds were identified by clicking and dragging across the waveform using the selection tool in Audacity. Next,  the selected cough sound was duplicated using the ``Duplicate'' option in the ``Edit'' menu, creating a new track consisting solely of the chosen cough sound. The duplicated sound was exported as a separate WAV file through the ``Export'' option in the ``File'' menu, allowing for individual labeling and analysis to preserve each cough sound as a separate entity. This process was repeated for each cough within the audio file, ensuring that all cough sounds were extracted and labeled accurately.

All the files with cough sounds were stored in the folder labeled ``cough''.
We made another folder with the label ``No cough'' for audio recordings without any cough sounds. Inside this folder, we put all kinds of background noises we hear daily, like typing on a keyboard, music playing, or people talking in the background. We did this because, in real situations, many sounds often happen simultaneously, and we want our machine-learning model to be good at differentiating a cough from other noises or sounds. The  ``No cough'' folder also allows us to check if machine learning models mistakenly consider these noises as coughs. So, it is an integral part of our research to ensure our model can do its job well in the real world, where all sorts of sounds are around.

After labeling,  we had 223 and 3500 cough and non-cough samples, respectively. Among the 223 cough samples, 139 were collected during non-walking activities (i.e., sitting and standing) and 84 during walking activity.

\subsubsection{Data Augmentation}

Augmentation enhances the machine learning model's resistance to input data fluctuations and its generalization ability to new data. We used the following two data augmentation \cite{liaqat2021coughwatch} techniques to augment the cough samples obtained during the study using Python's Numpy library.

\begin{itemize}
    \item Noise Augmentation: The noise augmentation function takes a scale factor and an audio signal input. It generates random white Gaussian noise using the \textit{np.random.normal} function, with the scale factor defining the standard deviation of the noise. It supplements the original audio signal with the noise and returns the augmented signal. We augmented cough data using this method and  the scale factor for adding  Gaussian noise to the cough data was set to 0.01.
    \item Interpolate Augmentation: The \textit{interpolate$\_$augmentation} function takes an audio signal as its input. It begins by discarding every second audio sample from the input signal, creating two new arrays: audio$\_$even containing even-indexed samples and audio$\_$odd containing odd-indexed samples.
Subsequently, the function uses the \textit{np.interp} function to perform linear interpolation on the dropped samples. For audio$\_$even, it interpolates the missing samples at even indices using the original indices, while for audio$\_$odd, it interpolates the missing samples at odd indices using the original indices. In simpler terms, the function takes the audio signal, removes every other sound, and then figures out what should go in the places where sounds were removed by connecting the remaining sounds together in a smooth and continuous way.
The result is two augmented audio signals, audio$\_$even and audio$\_$odd, which have the same length as the original audio signal but include interpolated values for the dropped samples.  Again we have augmented the cough data with this method and with a scale factor of 0.05
 
\end{itemize}

After augmenting the data, the cough samples increased to 892.

\subsection{Feature Extraction}

Feature extraction is necessary for audio processing applications, including cough detection. Features are extracted from the raw audio data to differentiate cough from other sounds or background noise.
The feature extraction technique often transforms raw audio data into a more helpful representation with significant signal features. This is accomplished by selecting the appropriate signal processing techniques and extracting relevant elements to distinguish coughs from other sounds.
 The two used features are the Mel-frequency cepstral coefficients (MFCCs) and the Mel-spectrogram.
 
\noindent{\bf MFCC:} It is a widely used feature extraction method for speech and audio processing. It mimics the human auditory system's ability to capture spectral information. The steps involved in MFCC extraction include segmenting the signal into small frames, applying window functions, performing the Fourier transform to obtain power spectrum, utilizing a Mel-filterbank to extract Mel-frequency components, applying logarithmic scaling, and finally subjecting the components to Discrete Cosine Transform (DCT) to produce the MFCC coefficients. These coefficients capture essential spectral characteristics and can be used for speech recognition and audio classification tasks.
 
\begin{figure}[ht]
     \centering
     \includegraphics[width=1\linewidth]{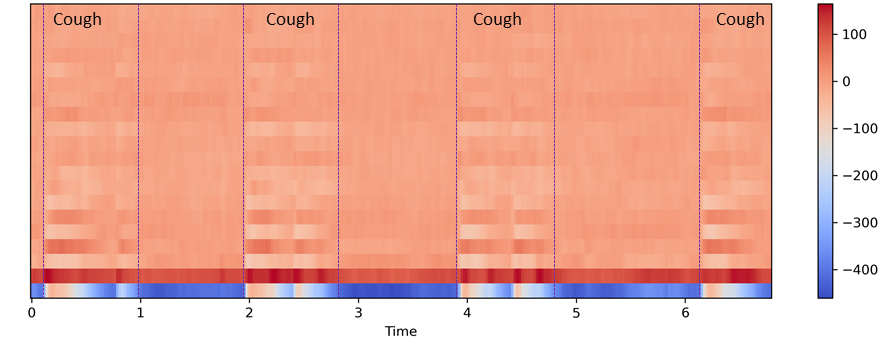}
     \caption{MFCC of an audio signal. [Best viewed in color]}
     \label{mfcc}
\end{figure}

In the MFCC figure in Figure \ref{mfcc}, the x-axis represents the audio signal frames, while the y-axis represents the MFCC coefficients. Each coefficient corresponds to a specific sound characteristic, such as pitch, timbre, or tone. The plot shows how these coefficients vary over time, providing valuable insights into the sound's qualities. This visualization helps to understand the temporal changes and patterns within the sound signal, aiding in sound classification, speech recognition, and audio analysis.
 
 \begin{figure}[h]
     \centering
\includegraphics[width=1\linewidth]{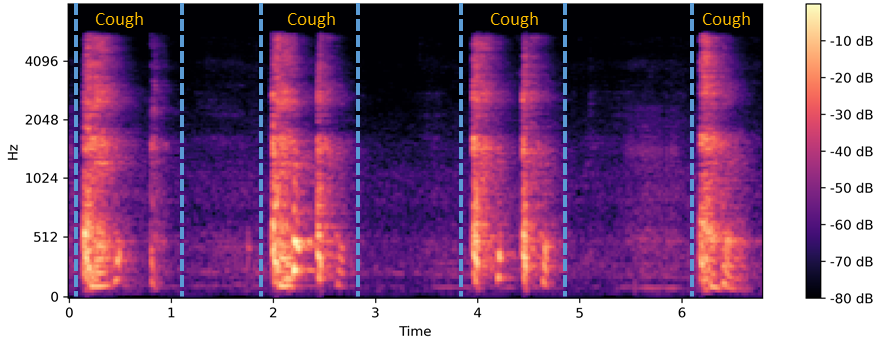}
     \caption{Mel Spectogram of an audio signal.[Best viewed in color]} 
     \label{mel}
 \end{figure}
 
\noindent{\bf Mel Spectogram:}
The Mel-spectrogram represents the power spectrum of a cough signal in the Mel-frequency domain. It involves segmenting the input signal into short frames, applying window functions to minimize spectral leakage, performing the Fourier transform on each frame to obtain the power spectrum, and applying a Mel-filterbank to create the Mel-spectrogram. The Mel-filterbank captures energy within specific frequency ranges based on the Mel scale. The resulting Mel-spectrogram visually displays the energy distribution across frequencies and time, providing valuable insights into the spectral characteristics of the cough signal.
 
The x-axis shows time, and the y-axis shows the frequency spectrum of a Mel-spectrogram in Figure \ref{mel}. The color's intensity corresponds to the strength of the sound wave at that particular frequency and instant. The plot shows that the frequency content of the sound signal varies over time, which reveals details about the sound's spectral properties.

We have used Librosa\cite{librosa}, an open-source Python library for feature extraction, and 40 (20 MFCCs and 20 from Mel spectrogram) feature values were extracted from each audio file.

\begin{table}[!h]
\centering
\caption{Architecture of 1D CNN Model.}
\resizebox{\columnwidth}{!}{%
\begin{tabular}{@{}lll@{}}
\toprule
Layer (type)                      & Output Shape    & Param \\ \midrule
conv1d\_4 (Conv1D)                & (None, 78, 128) & 512   \\
max\_pooling1d\_4 (MaxPooling 1D) & (None, 39, 128) & 0     \\
conv1d\_5 (Conv1D)                & (None, 37, 64)  & 24640 \\ 
max\_pooling1d\_5 (MaxPooling 1D) & (None, 18, 64)  & 0     \\
conv1d\_6 (Conv1D)                & (None, 16, 32)  & 6176  \\
max\_pooling1d\_6 (MaxPooling 1D) & (None, 8, 32)   & 0     \\
conv1d\_7 (Conv1D)                & (None, 6, 16)   & 1552  \\
max\_pooling1d\_7 (MaxPooling 1D) & (None, 3, 16)   & 0     \\
flatten\_1 (Flatten               & (None, 48)      & 0     \\
dropout\_2 (Dropout)              & (None, 48)      & 0     \\
dense\_2 (Dense)                  & (None, 64)      & 3136  \\
dropout\_3 (Dropout)              & (None, 64)      & 0     \\
dense\_3 (Dense)                  & (None, 32)      & 2080  \\
dropout\_4 (Dropout)              & (None, 32)      & 0     \\
dense\_4 (Dense)                  & (None, 1)       & 33  \\ \bottomrule
\end{tabular}
}
\label{tab:model1}
\end{table}

\subsection{Classification model}
We used 1D CNN (Convolutional Neural Network) for cough detection.
A time-series audio signal of 1-2 seconds is the model's input, and its output is a binary classification indicating whether the signal is cough or non-cough.

Table~\ref{tab:model1} provides an overview of the model's architecture, and the following is its description. It consists of an input layer that takes a two-dimensional array of features and time steps. It has four convolutional layers with varying filter sizes and ReLU activation functions. Max pooling layers are used after each convolutional layer to downsample the feature maps. The final max pooling layer output is flattened into a one-dimensional time vector. Two dropout layers are added to reduce overfitting. The flattened output is passed through two dense layers with ReLU activation functions. The output layer is dense with a sigmoid activation function for binary classification. The model is compiled with binary cross-entropy loss and the Adam optimizer. Accuracy is used as the evaluation metric, and early stopping is implemented to halt training if the validation loss does not improve for 100 epochs.

 \begin{table}[!t]

\caption{Accuracy, F1, Precision and Recall of different models during non-walking activities (i.e., sitting and standing)}
\centering
\resizebox{\columnwidth}{!}{%
\begin{tabular}{@{}lrrrr@{}}
\toprule
Models        & Accuracy & F1 Score & Recall & Precision \\ \midrule
Coughwatch AO & 0.8945   & 0.7596  & 0.7037 & 0.8252    \\
Random Forest & 0.9581   & 0.8877  & 0.8943 & 0.8811    \\
1D CNN        & 0.9849   & 0.9645  & 0.9583 & 0.9708    \\ \bottomrule
\end{tabular}
}
\label{w1}
\end{table}

\begin{table}[!t]
\centering
\caption{Accuracy, F1, Precision and Recall of different models during walking}
\resizebox{\columnwidth}{!}{%
\begin{tabular}{@{}lrrrr@{}}
\toprule
Models        & Accuracy & F1 Score & Recall & Precision \\ \midrule
Coughwatch AO & 0.8556   & 0.8324 & 0.8043 & 0.8625    \\
Random Forest & 0.9536   & 0.8460 & 0.8495  & 0.8426    \\
1D CNN        & 0.9823   & 0.8590 & 0.8482 & 0.8701    \\ \bottomrule
\end{tabular}
}
\label{w}
\end{table}

 \begin{figure}[]
     \centering
\includegraphics[width=1\linewidth]{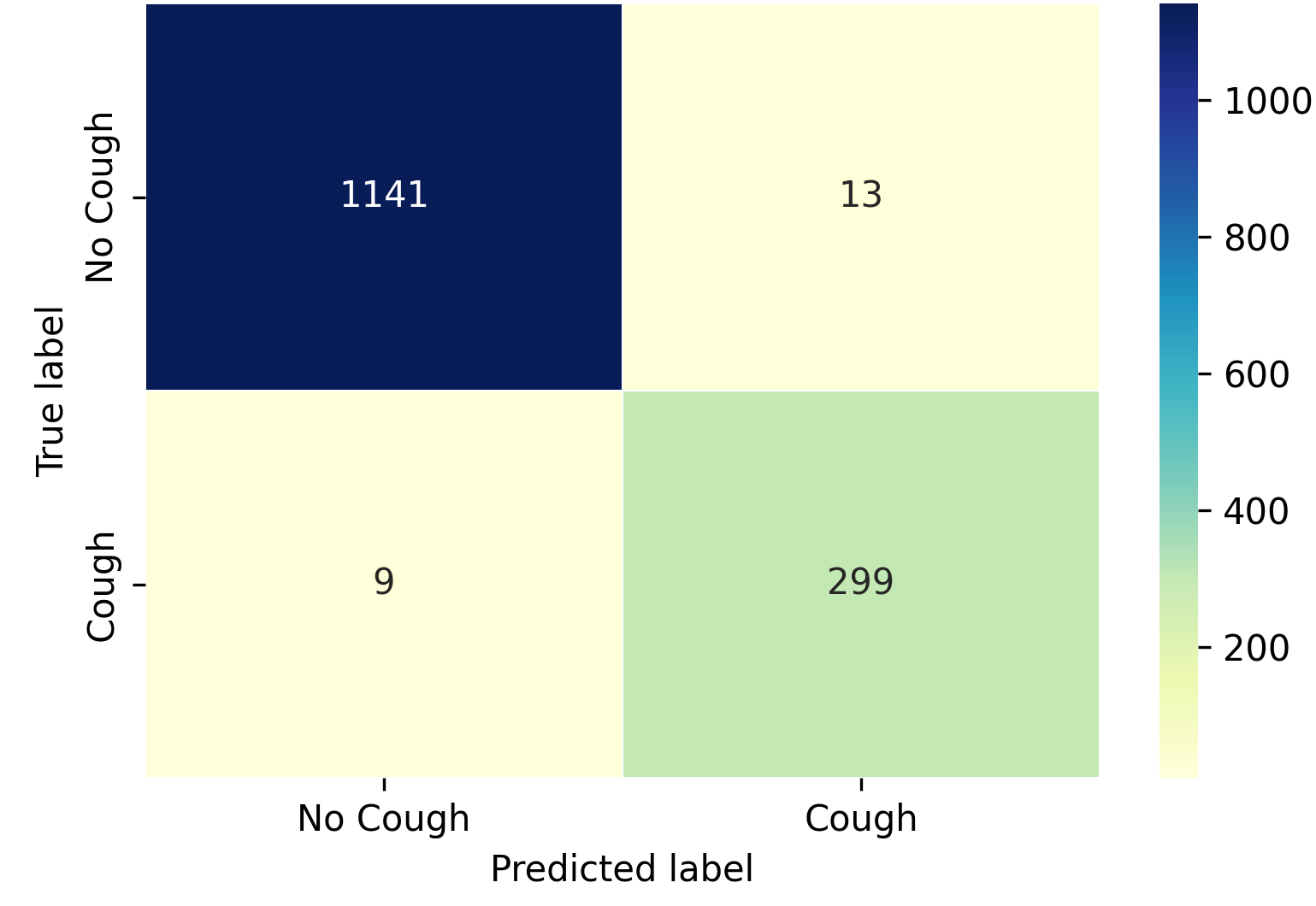}
     \caption{Confusion matrix of 1D CNN for cough classification during
     non-walking activities (i.e., sitting and standing).
     }
     \label{vec}
 \end{figure}

\begin{figure}[]
     \centering
\includegraphics[width=1\linewidth]{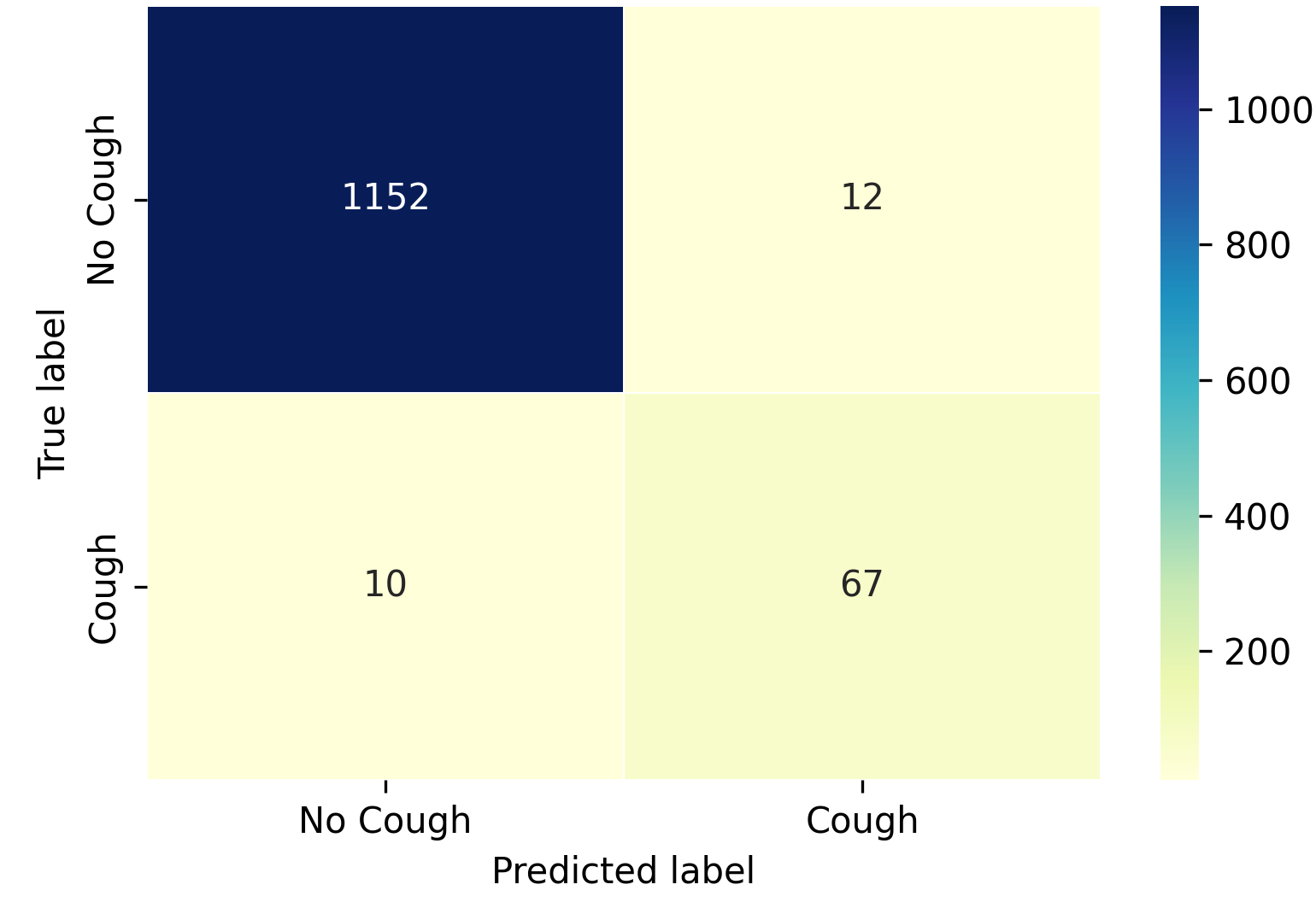}
     \caption{Confusion matrix of 1D CNN for cough classification during
     walking activity.
     }
     \label{vec1}
 \end{figure}

\section{Results}

Tables~\ref{w1} and \ref{w} report classification accuracy, F1 score, Precision, and Recall during non-walking (i.e., sitting and standing) an walking activities, respectively.
 The table reports the classification results with Coughwatch AO~\cite{liaqat2021coughwatch}, Random Forest, and 1D-CNN models.

For Non-walking activities, the 1D CNN model achieved the highest accuracy (0.9849) and F1 score (0.9645) of all models, outperforming Coughwatch AO and Random Forest.
Similarly, the 1D CNN model outperformed the other models in the cough detection while walking activity with an accuracy of 0.9823 and an F1 score of 0.8590. Although Random Forest had relatively high accuracy, the 1D CNN model showed superior accuracy, demonstrating its ability to effectively detect walking activity. In both cases, the 1D CNN model consistently showed the best overall performance, demonstrating its versatility and efficiency in classifying various features. 

Figures \ref{vec} and \ref{vec1} show the confusion matrix of the 1D CNN model for cough classification while non-walking (i.e., sitting and standing) and walking.
The accuracy achieved while non-walking is 98.49$\%$ and 98.23$\%$ while walking without overfitting, which suggests that the model has learned to differentiate between cough sounds and non-cough sounds quite well during both activities, suggesting that watches are pretty much capable of detecting cough while movement also. This level of accuracy could be considered high and may indicate that the model is ready for deployment in a real-world scenario.

\begin{figure}[h]
     \centering
\includegraphics[width=1\linewidth]{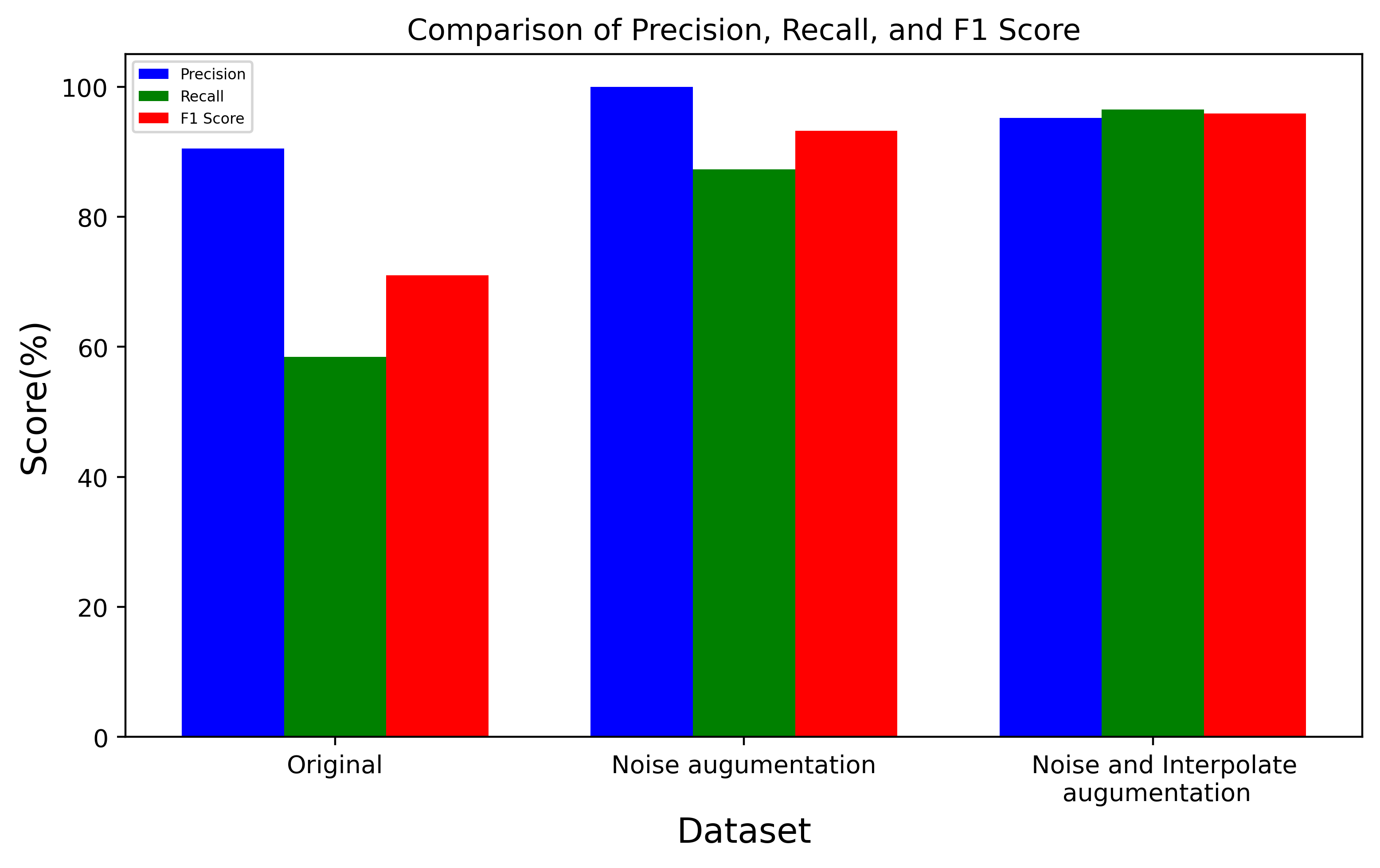}
     \caption{Precision, Recall, and F1-score with and without data augmentation during non-walking activities.}
     \label{mec}
 \end{figure}
\begin{figure}[!t]
     \centering
\includegraphics[width=1\linewidth]{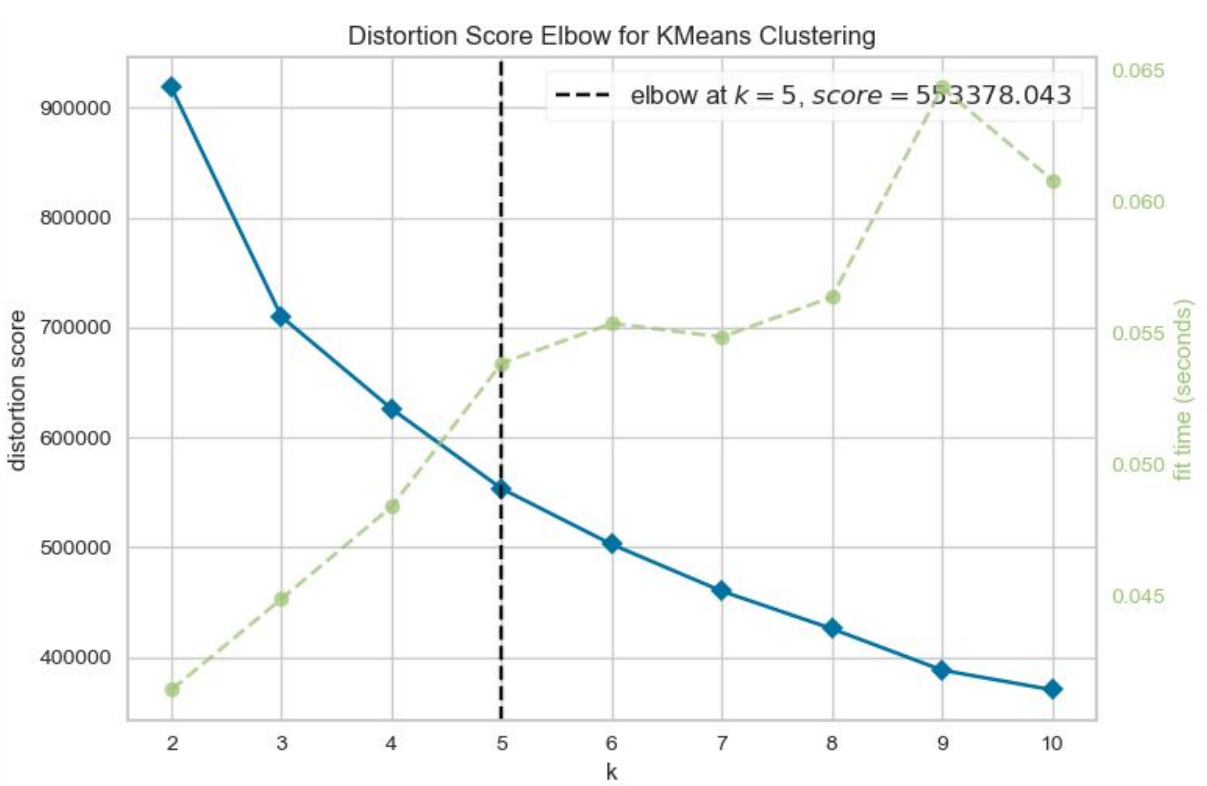}
     \caption{Elbow plot for clustering.}
     \label{elb}
 \end{figure}

\begin{figure*}[!t]
     \centering
\includegraphics[width=1\linewidth]{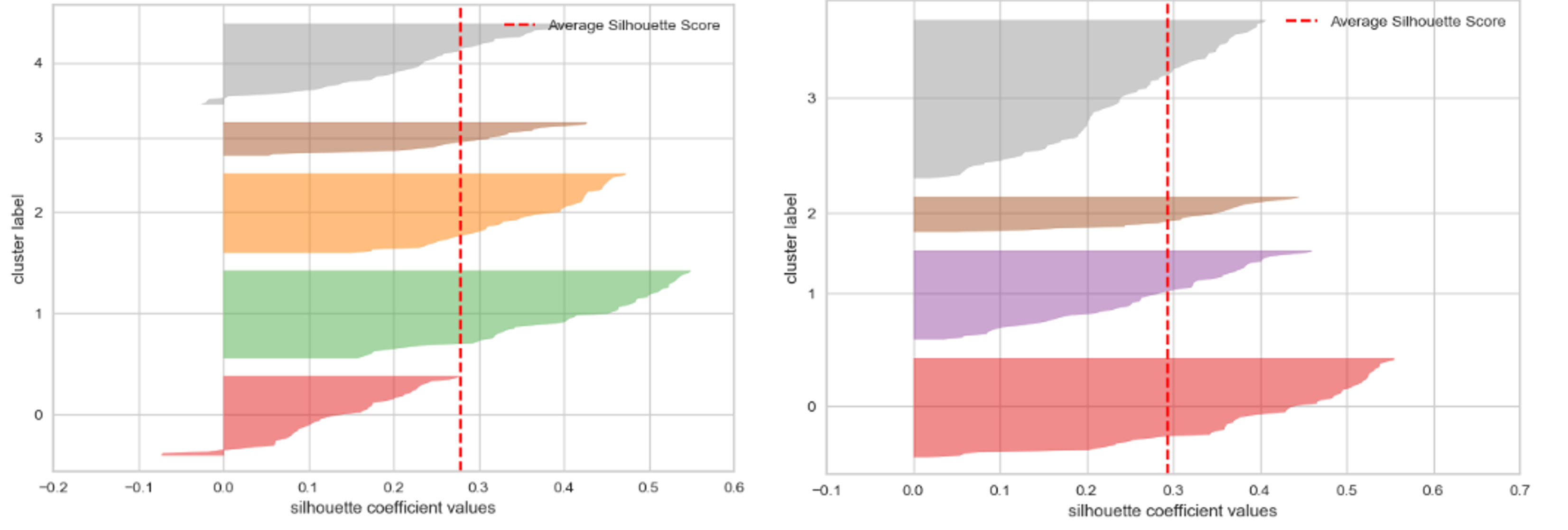}
     \caption{Comparison of average Silhouette scores with different number of clusters. [{\bf Best viewed in color}] }
     \label{sil}
 \end{figure*}
\emph{\bf Effect of Data Augmentation:}
We also evaluated the impact of data augmentation on different metrics such as Precision, Recall, and F1-score. 
Figure~\ref{mec} shows the results of all three metrics with original, noise augmented, and a mix of noise and interpolated datasets.  The plot shows that the data augmentations significantly increase precision, recall and F1 score.

\emph{\bf Cough types:}
We also did a clustering analysis to identify distinct types of coughs within our collected dataset. To validate and refine the clustering results, two commonly used methods, namely the elbow method and silhouette analysis, were employed to identify the number of clusters.

The elbow method~\ref{elb} revealed an initial suggestion of 5 clusters based on the within-cluster sum of squares (WCSS) metric. However, silhouette analysis was conducted to ensure the clustering's robustness and accuracy. Silhouette analysis calculates the silhouette coefficient for each data point, which measures the separation and cohesion within and between clusters. A higher silhouette coefficient indicates better clustering, with data points being more similar to their cluster than to other clusters. After performing silhouette analysis for various numbers of clusters, as shown in Figure~\ref{sil}, it was observed that the average silhouette scores were highest for 4 clusters. This number indicated better cluster separation and cohesion, implying distinct and well-defined cough patterns.

\begin{figure*}[]
     \centering
\includegraphics[width=1\linewidth]{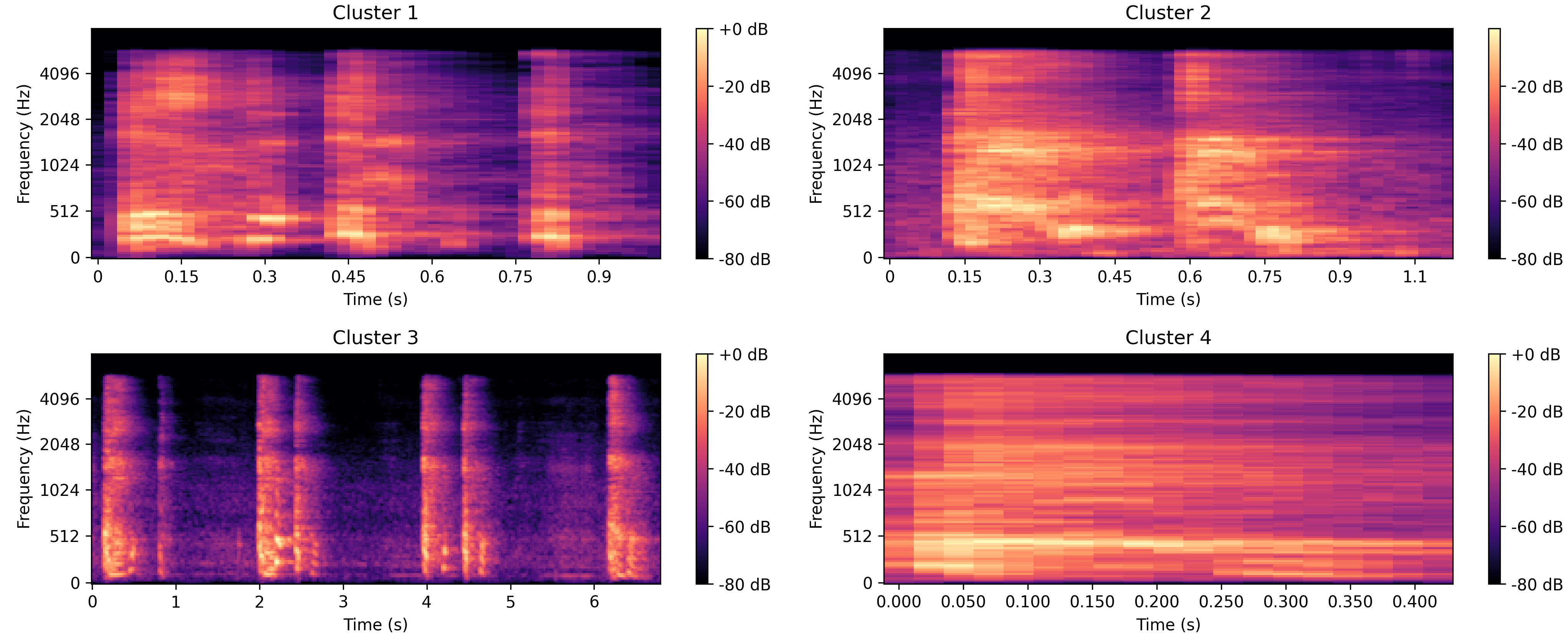}
     \caption{Mel Spectograms of four different cough clusters. [{\bf Best viewed in color}] }
     \label{c1}
 \end{figure*}

Therefore, based on the combined insights from the elbow method and silhouette analysis, four clusters were determined as the final number of cough types in the dataset. This decision ensured a meaningful and distinct grouping of cough samples within each cluster. Figure~\ref{c1} shows the Mel Spectograms of these four different cough clusters and 
Table~\ref{count} shows the cough frequency of each cluster.

The four categorized cough clusters have distinctive features. By analyzing the mel spectrogram and hearing the audio, the first cluster represents single cough events, likely indicative of standard or mild coughing. In the second cluster, double coughs suggest more frequent or intense coughing episodes. The third cluster consists of triple coughs, indicating a pronounced and prolonged coughing pattern. Lastly, the fourth cluster encompasses sequences of four coughs, potentially representing severe or intense coughing bouts. This clustering not only helps differentiate various cough intensities but also holds significant potential for healthcare applications, aiding in the monitoring and diagnosing of respiratory conditions by tracking changes in cough patterns over time.
 
Applying the elbow method and silhouette analysis strengthened the validity and interpretation of the clustering results. The elbow method initially estimated the number of clusters, while silhouette analysis offered a more comprehensive evaluation of the clustering quality. This comprehensive approach ensured a more accurate and insightful characterization of the cough samples, enhancing the understanding of different cough types and their distinct acoustic features within the audio dataset.
\begin{table}[!h]
\centering
\caption{Cough frequency of different clusters}
\begin{tabular}{@{}lrrrr@{}}
\toprule
Cluster number & 1  & 2  & 3  & 4  \\ \midrule
 Cough samples & 71 & 63 & 46 & 53\\ \bottomrule
\end{tabular}
\label{count}
\centering
\end{table}

\section{Discussion}
The results presented in this research paper demonstrate the significant potential of using smartwatches and advanced machine learning techniques for cough detection and classification. The implications of this work are far-reaching and have important implications for both healthcare professionals and individuals concerned about their respiratory health.

Firstly, the high accuracy achieved in cough detection (98.49$\%$) while non-walking and 98.23$\%$ while walking using the 1D CNN model is a promising indicator of the model's effectiveness in distinguishing cough sounds from other noises. The results suggest that the smartwatch application developed in this study can be reliable for continuously monitoring cough patterns, especially in individuals with respiratory health concerns.

Moreover, the clustering analysis identified four distinct types of coughs within the dataset. This discovery is significant as it implies that not all coughs are the same and may be associated with different underlying health conditions or behaviors. Differentiating between these cough types could potentially aid in early disease detection and provide valuable insights into an individual's respiratory health.

The practical applications of this research extend beyond chronic lung diseases. Cough monitoring can be crucial in identifying and tracking infectious diseases like COVID-19, as coughing is a prominent symptom associated with the illness. Establishing a baseline for respiratory health in healthy individuals and detecting deviations from this baseline can be invaluable for early disease detection and implementing public health measures.

\section{Conclusion and Future work}

We utilized a 1D CNN model for cough detection and achieved an impressive accuracy of 98.49\% while sitting and 98.23\%. The model exhibited exceptional performance in effectively distinguishing between cough sounds and non-cough sounds. The high accuracy indicates that the model is well-suited for real-world applications in cough detection. Furthermore, we applied clustering techniques to identify distinct types of coughs within our dataset. The analysis revealed the presence of four different types of coughs, demonstrating that the smartwatch can be utilized to detect and classify coughs associated with specific issues. These findings highlight the potential of using wearable devices, such as smartwatches, in detecting and categorizing various types of coughs.

 Future work could focus on refining the algorithms, conducting larger-scale studies, and addressing the practical challenges associated with wearable devices to enhance further the applicability of this technology in healthcare and public health settings.

\bibliographystyle{IEEEtran}
\bibliography{IEEEabrv,citationfile}

\end{document}